# Visualizing the Doppler Effect

Marcos H. Giménez[*], Ana Vidaurre, Jaime Riera, and Juan A. Monsoriu

*Departamento de Física Aplicada, Universidad Politécnica de Valencia,
E-46022 Valencia, Spain*

**ABSTRACT**

The development of Information and Communication Technologies suggests some spectacular changes in the methods used for teaching scientific subjects. Nowadays, the development of software and hardware makes it possible to simulate processes as close to reality as we want. However, when we are trying to explain some complex physical processes, it is better to simplify the problem under study using simplified pictures of the total process by eliminating some elements that make it difficult to understand this process. In this work we focus our attention on the Doppler effect which requires the space-time visualization that is very difficult to obtain using the traditional teaching resources. We have designed digital simulations as a complement of the theoretical explanation in order to help students understand this phenomenon.

---

[*] Electronic mail: mhgimene@fis.upv.es

# 1. INTRODUCTION

In recent years, Information and Communication technologies (hereafter ICTs) are being used in many ways in teaching science. They permit interactivity, in agreement with the constructivism pedagogical principles. The students have to construct their own body of knowledge so the teacher guides the process providing the student with the necessary tools to get it [1].

Different works have shown the advantages of digital simulation in the learning process [2, 3]. Actually a lot of online multimedia material for teaching physics is available [4]. However, one should know that the pedagogical effectiveness is not directly related with the use of simulation but with its correct integration in the global instruction plan [5, 6]. As the ICTs are now in a developing process, research about the new possibilities they offer in the education field is required. Recently, our working group has developed several simulation programs designed in such a way that they underline the visualization of the physical problem [7, 8]. In this way, the teacher is able to obtain the major benefits of the computer-based instruction.

Learning based on models is more likely to aid comprehension than superficial memorization. Mental models play a very important role in activities related to the learning process. ICTs offer a good opportunity to develop these ideas through 3D animations immersed in virtual environments.

Wave motion, in general, and Doppler effect, in particular, require the simultaneous understanding of the spatial and time dependence [8]. Thus, the traditional teaching resources—blackboard, slides, transparencies— are not enough. If we want to show motion we need the real process or the videotape, previously recorded. Both techniques are very useful in the learning process because they allow to establish connections between theory and reality.

However, in some cases, it is not possible to eliminate some elements that make it difficult to understand the essential aspects of the problem under study.

This can be solved using digital animations, now possible in the ICTs context. This technique presents the following advantages:

- Similarly to the cinema or television, simulations produce the effect of a continuous representation in time. This is not possible with slides or transparencies.

- They appear free of noise and allow us to identify the different aspects we want to emphasize. In this sense simulations can be more efficient than the real process itself.

We have been working in this field designing non-interactive virtual animations that illustrate physical concepts as a complement of the theoretical explanation. In the present work we present the process we have followed in order to design digital simulations to help students in the understanding of the Doppler effect. These animations have been developed in a virtual environment using the program 3D Studio Max and formatted in digital video files.

## 2. DIGITAL ANIMATIONS OF THE DOPPLER EFFECT.

The goal of the simulations here presented is to visualize the Doppler effect. We consider a source of sinusoidal wave motion, S, and the observer O. One of them, or both, is in motion so that the frequency measured by the observer, $v'$, differs from that generated by the source, $v$. The relation between them is given by the well-known equation:

$$v' = v \frac{v - v_O}{v - v_S}, \qquad (1)$$

where v, $v_S$, and $v_O$ are the propagation speed, the source speed and the observer speed, respectively.

As shown in Figure 1, the design of the simulations is based on the following characteristics:

- Each animation includes the top view where the wave source (dark color sphere) and one or more observers (light color spheres) in motion are shown. In many cases, the front view is also added.

- The wave fronts are represented as circumferences centered at the point where the wave was generated and the radius increases proportionally to the propagation speed. The dark ones represent the crests and the light ones the valleys.

- The front view shows the wave being generated by the vertically oscillating source, and affecting the observes. In this way the animation allows an easy comparison between the oscillation frequency of the source and the observers.

This scheme offers students different points of view that cover different concepts related with the Doppler effect. The simultaneous representation of the time and space dependence improves the understanding of the magnitudes involved: period, frequency, wavelength, and relative motion of the observer respect to the source. The different aspects can be analyzed one by one or globally. The qualitative representation can be compared with the mathematical expression given by equation (1).

Following the above scheme, we have developed animations in order to study particular situations related with the motion of the source, the observer, or both of them. These situations are:

- The source moves to the right with a lower speed than that of the propagation motion and the observers stay at rest (Fig. 1). The left-side observer measures a lower frequency than that generated in the source. The right-side observer, on the contrary, measures a higher frequency than the source does. In addition to the Doppler effect, we can use this example to demonstrate the relationship between period and wavelength.

- The source is at rest and the observer moves more slowly than the propagation speed (Fig. 2). The left-side observer, who moves away from the source, oscillates with a frequency lower than that corresponding to the source. The right-side observer, who approaches to the source, oscillates with a higher frequency than that of the source. In this example the importance of the relative motion can be clearly appreciated.

- The source stays at rest and the observer travels at the same speed as the generated waves (Fig. 3). Then the observer does not perceive the wave motion, and the measured frequency is zero according to eq. (1).

- The source stays at rest while the observer moves with a higher speed than the propagation wave (Fig. 3). The observer oscillates under the action of the generated wave front in inverse order; in other words, the frequency is negative, as predicted by eq. (1).

The proposed scheme can also be applied to other processes related to source motion:

- The source moves with the same speed as the generated waves (Fig. 7). The wave fronts are all superimposed just in front of the source generating the wave barrier. Typical examples of this phenomenon are sound waves generated by airplanes (sound barrier) and water superficial waves generated by ships.

- The source moves with higher speed than the generated waves (Fig. 8). A shock wave is generated as the envelope of the wave front.

## 3. CONCLUSIONS

The understanding of wave motion requires space-time visualization, which is very difficult to obtain using the traditional teaching resources —blackboard, transparencies, slides...— because these techniques only allow to show static images. Some other techniques such as experiments (real or in video) mix different aspects involved in the process and it is difficult to show the essential aspects that the teacher wants to transmit. This problem can be solved with digital animations that provide a continuous time representation of the physical phenomenon and, on the other hand allow us to "clean" the process showing only the aspects we want to emphasize in order to get a better understanding of it. In the present work, we have presented the design of digital animations for the explanation of the Doppler effect. The scheme consists basically of showing the top and front views of the wave fronts generated by the source. The wave front travels in a homogeneous medium and generates oscillations in the observers. This scheme makes it easier the understanding of the Doppler effect and other related problems like sound barrier or shock waves.

**AKNOWLEDGEMENTS**

This work was supported by Plan de Incentivo a la Investigación form Universidad Politécnica de Valencia (grant 20020632), Spain.

**REFERENCES**

[1] T. Duffy and D. Jonassen, Constructivism and the technology of instruction, Lawrence Erlbaum Associates, Hillsdale, New Jersey (1992).

[2] D.J. Grayson and L.C. McDermott, Use of the computer for research on student thinking in physics. Am. J. Phys. 64, 557-565 (1996).

[3] F. Esquembre, Computers in physics education. Comput. Phys. Commun 147, 13-18 (2002).

[4] S. Altherr, A. Wagner, B. Eckert, and H.J. Jodl, Multimedia material for teaching physics (search, evaluation and examples), Eur. J. Phys. 25, 7-14 (2004).

[5] D. Hestenes, Who needs physics education research?, Am. J. Phys. 66, 465-467 (1998).

[6] R.N. Steinberg, Computers in teaching science: To simulate or not to simulate?. Am. J. Phys. 68, S37-S41 (2000).

[7] A. Vidaurre, J. Riera, M.H. Giménez and J.A. Monsoriu, Contribution of digital simulation in visualizing physics processes. Comput. Appl. Eng. Educ. 10, 45-49 (2002).

[8] J. Riera J, M.H. Giménez, A. Vidaurre, and J.A. Monsoriu, Digital simulation of wave motion. Comput. Appl. Eng. Educ. 10, 161-166 (2002).

**FIGURE CAPTIONS**

**Figure 1.** Two frames (separated by a source oscillation period) of a digital animation of the Doppler effect where the source moves to the right with a lower speed than that of the propagation and the observers stay at rest.

**Figure 2.** Two frames (separated by a source oscillation period) of a digital animation of the Doppler effect where the source is at rest and the observer moves more slowly than the propagation speed.

**Figure 3.** Two frames (separated by a source oscillation period) of a digital animation of the Doppler effect where the source and the observer are both in motion.

**Figure 4.** Two frames (separated by a source oscillation period) of a digital animation of the Doppler effect where the source is moving slower than the propagation speed of the generated waves while the observer stays at rest.

**Figure 5.** Two frames (separated by a source oscillation period) of a digital animation of the Doppler effect where the source stays at rest and the observer travels at the same speed as the generated waves.

**Figure 6.** Two frames (separated by a source oscillation period) of a digital animation of the Doppler effect where the source stays at rest while observer moves with a high spped than the propagation wave.

**Figure 7.** Two frames (separated by a source oscillation period) of a digital animation related with source motion (wave barrier) where the source moves with the same speed as the generated waves.

**Figure 8.** Two frames (separated by a source oscillation period) of a digital animation related with source motion (shock wave) where the source moves with higher speed than the generated waves.

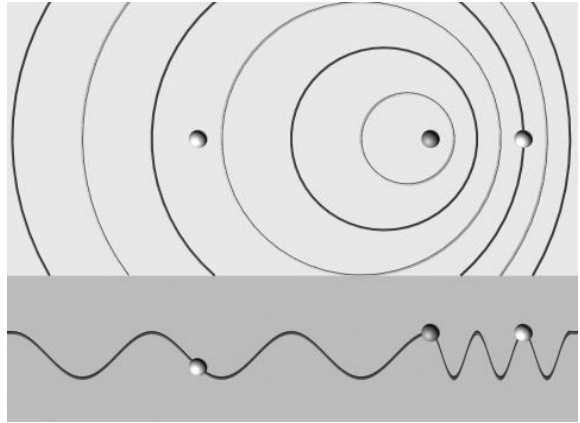
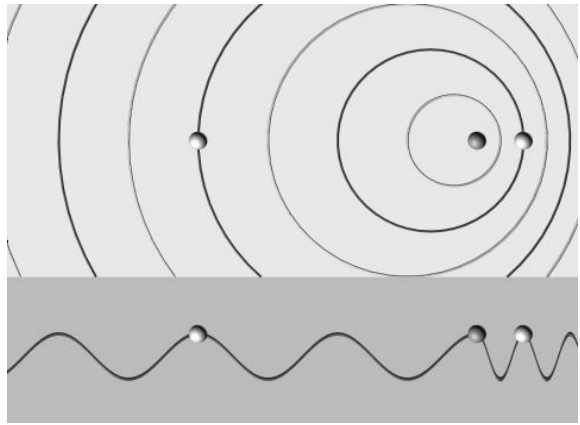

Figure 1
M.H. Giménez *et al.*

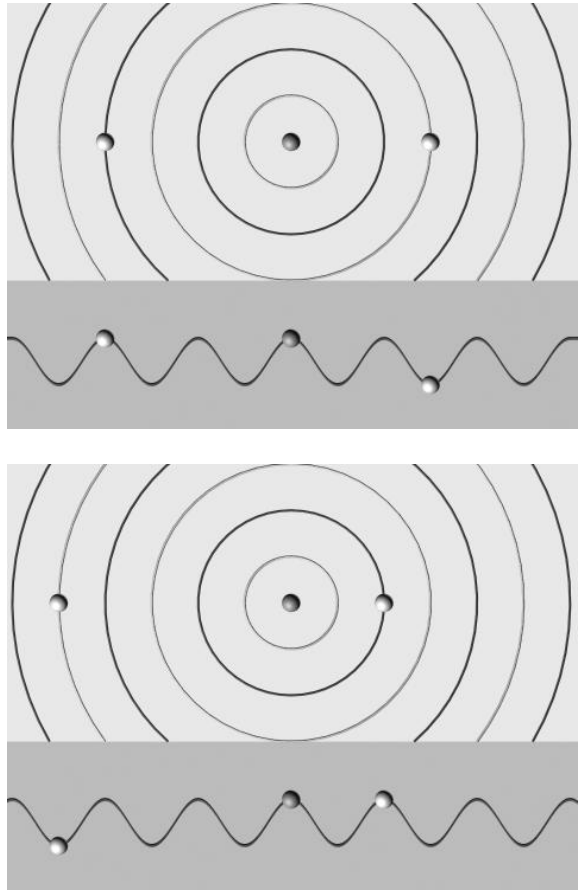

Figure 2
M.H. Giménez *et al*.

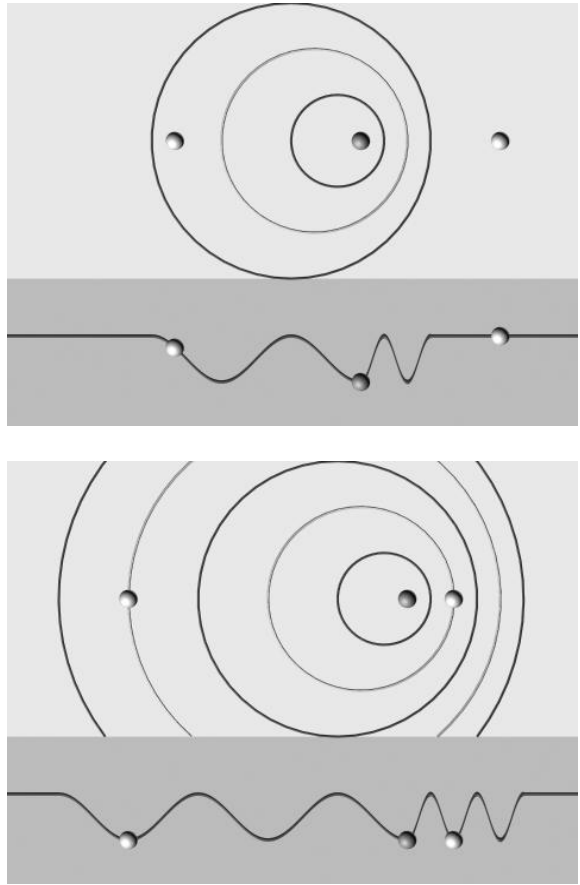

Figure 3
M.H. Giménez *et al*.

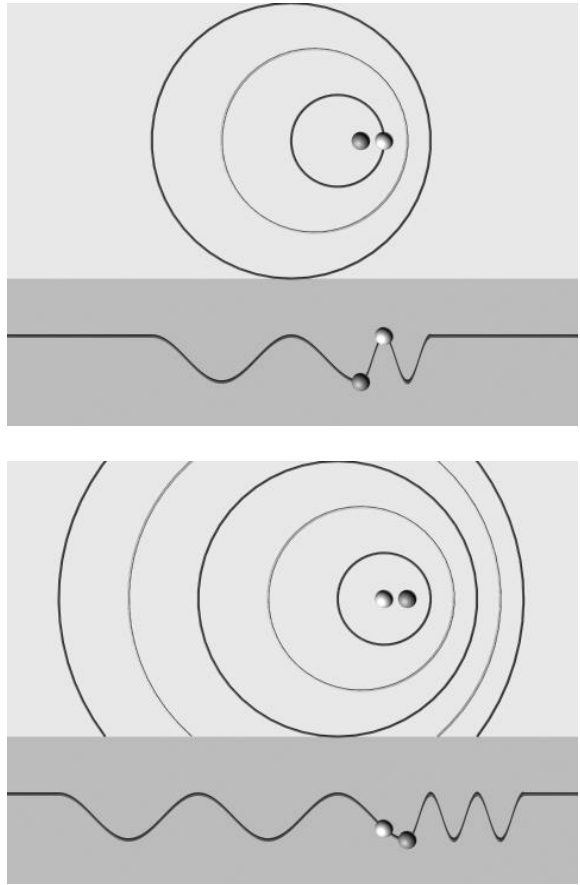

Figure 4
M.H. Giménez *et al*.

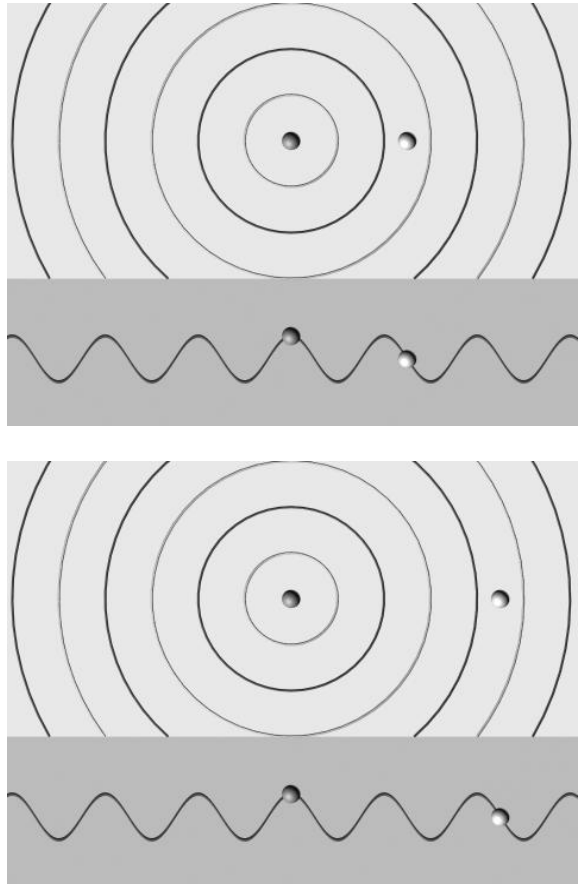

Figure 5
M.H. Giménez *et al*.

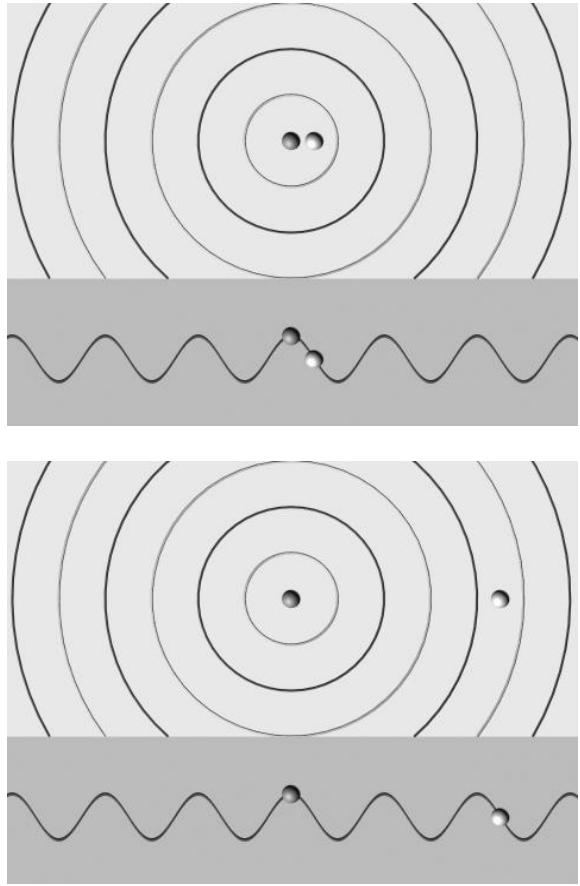

Figure 6
M.H. Giménez *et al.*

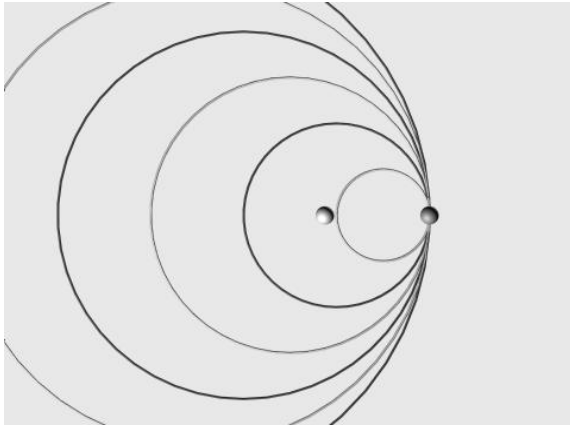

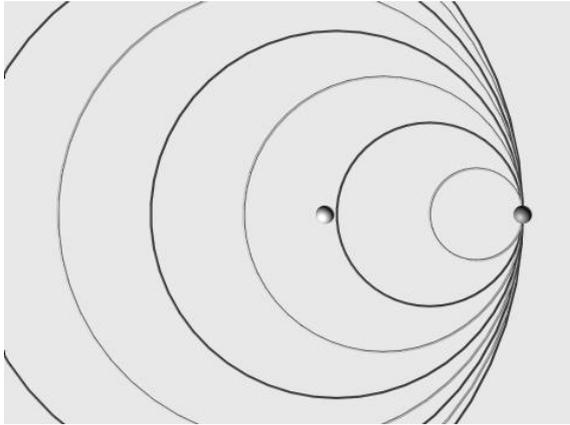

Figure 7
M.H. Giménez *et al*.

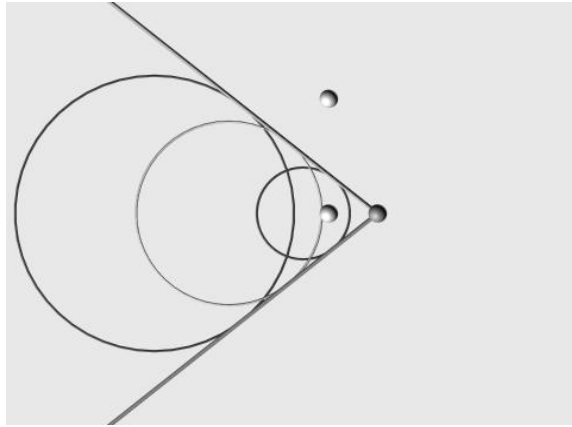

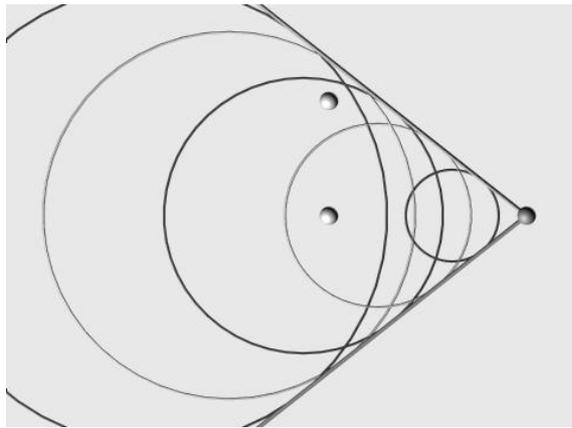

Figure 8
M.H. Giménez *et al.*